\documentclass[10pt]{article}  
\usepackage{graphicx}
\usepackage{amsmath, amssymb}
\usepackage[usenames,dvipsnames]{color}
\input epsf
\parskip=\medskipamount
   \def\CaL{{\cal L}}

\def\al{\alpha}
\def\be{\beta}
   \def\Ga{\Gamma}

\def\om{\omega}   
      
\def\IB{\relax{\rm l\kern-.18 em B}}
\def\IC{\relax{\rm l\kern-.50 em C}}
\def\IE{\relax{\rm l\kern-.12 em E}}
\def\IK{\relax{\rm l\kern-.18 em K}}
\def\IL{\relax{\rm I\kern-.18 em L}}
\def\IN{\relax{\rm I\kern-.18 em N}}
\def\IR{\relax{\rm I\kern-.18 em R}}

\def\Ker{\mathop{\rm Ker}\nolimits}

\def\smallonehalf{\frac{{}_1}{{}^2}}
\def\wh{\widehat}

\def\frac#1#2{{#1\over #2}}
\def\fracpd#1#2{\frac{\partial #1}{\partial #2}}

\def\ptos{\leaders\hbox to 2mm{\hfil{.}\hfil}\hfill}
\def\\{\hfill\break}

\def\<#1>{\langle#1\rangle}

\font\tenfrak=eufm10  \font\sevenfrak=eufm7  \font\fivefrak=eufm5
\newfam\frakfam
\textfont\frakfam=\tenfrak\scriptfont\frakfam=\sevenfrak
\scriptscriptfont\frakfam=\fivefrak

\font\tengoth=eufm10 scaled\magstep1 \font\sevengoth=eufm7
\font\fivegoth=eufm5
\newfam\gothfam
\textfont\gothfam=\tengoth\scriptfont\gothfam=\sevengoth
   \scriptscriptfont\gothfam=\fivegoth

\newtheorem{proposition}{Proposition}

\begin{document}

\title{ Complex functions and geometric structures associated to the superintegrable Kepler-related  family of  systems endowed with generalized Runge-Lenz  integrals of motion }

\author{  
Manuel F. Ra\~nada  \\ [3pt]
{\sl Departamento de F\'{\i}sica Te\'orica and IUMA } \\
 {\sl Universidad de Zaragoza, 50009 Zaragoza, Spain}     }
\maketitle 

\begin{abstract} 

The existence of quasi-bi-Hamiltonian structures  for a two-dimensional superintegrable $(k_1,k_2,k_3)$-dependent Kepler-related problem is studied.
We make use of an approach that is related with the existence of some complex functions which satisfy  interesting Poisson bracket relations and that was previously applied to the standard Kepler problem as well as to some particular superintegrable systems as the Smorodinsky-Winternitz (SW) system,  the Tremblay-Turbiner-Winternitz (TTW) and  Post-Winternitz (PW) systems. 
We prove that these complex functions  are important for two reasons: 
first, they determine the integrals of motion, and second they determine the existence of some  geometric structures 
(in this particular case,  quasi-bi-Hamiltonian structures).

 All the results depend of three parameters ($k_1, k_2, k_3$) in such a way that in the particular case $k_1\ne 0$, $k_2=  k_3=  0$,  we recover the results of the original Kepler problem (previously studied in SIGMA 12,   010 (2016)). 
This paper can be considered as divided in two parts and every part present a different approach (different complex functions and different quasi-bi-Hamiltonian structures).

\end{abstract}

\begin{quote}

{\sl Running title:}{\enskip}
Quasi-bi-Hamiltonian structures  of a superintegrable Kepler-related  problem.


\end{quote}

\vfill
\footnoterule{\small
\begin{quote}
  {\tt E-mail: {mfran@unizar.es}  }
\end{quote}
}

\newpage


\section{Introduction}

In a recent paper the existence of quasi-bi-Hamiltonian structures for the 2-dimensional Kepler problem was studied \cite{CarRan16Sigma}. Now we present a similar study but for a  family of superintegrable Kepler-related systems. 

 In fact, the main purpose of this paper is to present a geometric study of the properties of a family of superintegrable Kepler-related systems depending on three parameters $k_i$, $i=1,2,3$. 
 We will prove that it admits  quasi-bi-Hamiltonian structures and, in order to arrive to this result, we will make use of an approach that is related with the existence of some complex functions which satisfy  interesting Poisson bracket relations.  
This formalism was previously applied to the study of, not only the standard Kepler problem  \cite{CarRan16Sigma}, but also to  other superintegrable two-dimensional systems as 
the nonlinear isotonic oscillator (SW system) \cite{Ran17Jmp} or the Tremblay-Turbiner-Winternitz (TTW) and the Post-Winternitz (PW) systems \cite{Ran17JPa}.

So we first recall some basic facts characterizing  superintegrability and quasi-bi-Hamiltoninan structures.

First,  it   is known the existence of   four  families of potentials whith separability in two different coordinate systems in the Euclidean plane and that they are, therefore, superintegrable with quadratic in the momenta constants of motion \cite{FrMaSmUW65} --\cite{MiPoWJPa13}. 
The two first potentials are related with the harmonic oscillator and they are not considered in this paper. 
The other two are the following  potentials 

\begin{enumerate}
\item[(K1)]  The following $(k_1,k_2,k_3)$-dependent Kepler-related potential  
\begin{equation}   
  V_{K1} =  \frac{k_1}{\sqrt{x^2 + y^2}}  +  \frac{k_2}{y^2}  +  \frac{k_3x}{y^2 \sqrt{x^2 + y^2}}
\end{equation}   
is separable in  (i) polar coordinates $(r,\phi)$ and  (ii) parabolic coordinates $(a,b)$. 

\item[(K2)]  The following $(k_1,k_2,k_3)$-dependent Kepler-related potential  
\begin{equation}   
V_{K2} =  \frac{k_1}{\sqrt{x^2 + y^2}}   
  + k_2 \frac{ \bigl[\sqrt{x^2 + y^2} + x\bigr]^{1/2} }{ \sqrt{x^2 + y^2}}   
  + k_3 \frac{ \bigl[\sqrt{x^2 + y^2} - x\bigr]^{1/2} }{ \sqrt{x^2 + y^2}}
\end{equation}   
 is separable in  (i) parabolic coordinates $(a,b)$ and  (ii) a second system of parabolic coordinares $(\alpha,\beta)$ obtained from $(a,b)$ by a rotation. 
\end{enumerate}

At this point we recall that the superintegrability of the rational harmonic oscillator (non-central harmonic oscillator with rational ratio of frequencies) 
$$
  H_{mn}  =  \frac{1}{2} \bigl( p_x^2 +  p_y^2 \bigr) + \frac12 \alpha_0^2\bigl( m^2 x^2 +  n^2 y^2 \bigr) \,,    
$$
can be proved making use of the complex functions $A_x$ and $A_y$ \cite{JauchHill40,{Perelomov}, {CarMarRan02}}, 
defined as
$$
 A_x = p_x + i\,m\,{\alpha_0} x   \,,{\quad}   A_y =p_y + i\,n\, {\alpha_0} y  \,, 
$$
that satisfy 
$$
 \frac{d}{d t}\,A_x  = \{A_x\,, H_{mn}\} = i\,m\,{\alpha_0} A_x \,,{\qquad}
 \frac{d}{d t}\,A_y  = \{A_y\,, H_{mn}\}  = i\,n\,{\alpha_0} A_y \,.  
$$
Then, the function $A_{xy}$ defined as $A_{xy} =(A_x)^n(A_y^*)^m$, is a constants of motion 
(the two real functions $ |A_{xx}|^2$ and $|A_{yy}|^2$ are just the two one-dimensional energies $E_x$ and $E_y$) 
and since it is a complex function, it determines not one but two real first integrals, Re$(A_{xy})$ and Im($A_{xy})$ (we have obtained four integrals but, since the system is two dimensional, only three of them can be independent).

The important point is that  this property (superintegrability  related with the existence of some complex functions satisfying certain Poisson brackets properties) is not just an exclusive characteristic of the harmonic oscillator $H_{mn}$. 
In fact, it has been recently proved that other superintegrable systems also admit a complex factorization for the additional constants of motion   (as the above mentioned SW nonlinear isotonic oscillator \cite{Ran17Jmp}, Tremblay-Turbiner-Winternitz (TTW) and  Post-Winternitz (PW) systems \cite{Ran17JPa} and also some particular systems defined in spaces with constant curvature \cite{Ran14JPa, Ran15PLa}).

Second, Suppose that the  phase space of a Hamiltonian system, that is,  the $2n$--dimensional cotangent bundle $T^*Q$ of the configuration space $Q$ endowed with the canonical symplectic structure $\om_0$, 
is equipped  with a second symplectic structures $\om_1\ne \om_0$. 
Then a  vector field $\Ga$ is said to be  bi--Hamiltonian  if it is Hamiltonian with respect to both structures, that is, 
\begin{equation}
  i(\Ga)\,\om_0 = d H_0 \,,\quad{\rm and}\quad i(\Ga)\,\om_1 = d H_1 \,. \label{biH}
\end{equation}
Hence, we have two distinct Hamiltonian formulations for the same  dynamical system 
(we note that in some cases $\om_1$ can be  a closed  but nonsymplectic 2--form). 
A consequence is that the pair $(\omega_0,\omega_1)$ determines a $(1,1)$ tensor field $R$ defined as
\begin{equation}
  \omega_1(X,Y) = \omega_0(RX,Y) \,,{\quad} \forall X,Y \in \mathfrak{X}(T^*Q) 
\end{equation}
in such a way that $R$  is $\Gamma$-invariant and  the eigenfunctions of $R$ are constants of motion. 
If $R$ has $n$ distinct eigenfunctions and in addition the Nijenhuis tensor $N_R$ of the tensor field $R$ vanishes, then  the system is Liouville integrable \cite{Fernandes94,  CrampSarTh00}. 
Bi-Hamiltonian system satisfying just (\ref{biH}) are usually called weak bi-Hamiltonian systems (in opposition to strong structures satisfying the Nijenhuis condition); for example, systems admitting canonoid transformations \cite{CarRanJmp88}  or non-symplectic symmetries \cite{CarMarRan02}, that are known to be   bi-Hamiltonian, can be just weak bi-Hamiltonian.
 
The point is that  bi-Hamiltonian structures are very interesting but, in most of cases,   difficult to be obtained. A consequence had been the convenience of introducing the related but weaker concept of  quasi-bi-Hamiltonian system  \cite{BrouCaboz96}--\cite{Blasz09}.

A Hamiltonian vector field $\Ga$ on  $(T^*Q,\omega_0)$ is called  quasi-bi-Hamiltonian if, in addition, it is quasi-Hamiltonian with respect to another symplectic structure  $\omega_1\ne\omega_0$.  That is, there exists a (nowhere-vanishing) function $\mu$ such that it satisfies the equation  $i(\mu\, \Ga)\,\omega_1 = dh$  for some function $h$  
(this function $h$ is a first integral of $\Ga$).
So we have 
\begin{equation}
  i(\Ga)\,\om_0 = d H_0 \,,\quad{\rm and}\quad i(\mu \Ga)\,\om_1 = d h \,. \label{qbiH}
\end{equation}

Next we summarize the contents of this paper.

First. 
We will study the Poisson bracket properties  of some particular complex functions  and then we will prove that the superintegrability of the $H_{K2}$ system is very related with the properties of these complex functions 
(we will prove the existence of two different approaches).

Second. 
We will prove that thse complex functions determine the existence of several (complex and real) quasi-bi-Hamiltonian structures. 

 All the results obtained in this paper depend of the three parameters ($k_1, k_2, k_3$) in such a way that in the particular case $k_1\ne 0$, $k_2=  k_3 =  0$,  we recover the results of the original Kepler problem \cite{CarRan16Sigma}.

 We must clearly advance that we will obtain structures  (wedge product of the differentials of complex functions) that do not satisfy the above mentioned Nijenhuis torsion condition (this was also true in the $k_2=  k_3 =  0$ Kepler case \cite{CarRan16Sigma}); so they are in fact weak quasi-bi-Hamiltonian structures (in opposition to strong structures satisfying the Nijenhuis condition).
Nevertheless, the purpose in this paper is not to prove the integrability of a system as consequence of a bi-Hamiltonian structure; in fact, we recall that  the multiple separability of $V_{K2}$ was known since \cite{FrMaSmUW65}.
 The main idea is that the superintegrable systems are systems endowed with interesting properties deserving be studied. Now,  in this paper,  we obtain several new properties all of them  related with the above mentioned complex functions.

\section{Hamiltonian $H_{K2}$.  Complex functions, Superintegrability,  and quasi-bi-Hamiltonian structures  } \label{Sec2}

In what follows we will study the second Kepler-related system making use of parabolic coordinates that we denote by $(a,b)$.  
First we recall that the two linear momenta and the angular momentum take the form 
$$
 P_1 = \frac{a p_a - b p_b}{a^2+b^2}    \ ,\ 
 P_2 = \frac{a p_b + b p_a}{a^2+b^2}    \ ,\ 
 J = a p_b - b p_a   \,,   \label{defJPxPy} 
$$
and also that if a natural  Euclidean Hamiltonian  $H=T+V$ takes the form  
$$
  H = \frac{1}{2m}\,\Bigl(\frac{p_a^2 + p_b^2}{a^2 + b^2}\Bigr)  + V(a,b)  \,,{\quad}
V(a,b) = \frac{F(a) + G(b)}{a^2 + b^2}   \,, 
$$
then it is Hamilton--Jacobi  separable and,  therefore, Liouville integrable  with the following quadratic function 
$$
 I_2 = (a p_b - b p_a) \Bigl(\frac{a p_b + b p_a}{a^2+b^2} \Bigr) + 
 2 \Bigl(\frac{a^2G(b)-b^2F(a)}{a^2+b^2}\Bigr)
 =  J P_2 +  2 \Bigl(\frac{a^2G(b)-b^2F(a)}{a^2+b^2}\Bigr)
$$
as the second constant of motion (the first one is the Hamiltonian itself).

 Now we consider the Hamiltonian  $H_{K2}$  of the Kepler-related superintegrable potential $V_{K2}$.  
It takes the following form when written in parabolic coordinates 
\begin{equation}   
  H_{K2}  =   ({\smallonehalf})\,\Bigl( \frac{p_a^2 + p_b^2}{a^2+b^2} \Bigr) +
    \Bigl[\,  \frac{k_1}{a^2+b^2} + \frac{k_2\, a }{a^2+b^2} +  \frac{k_3\, b}{a^2+b^2}  \,\Bigr]   \,. 
\end{equation}   
Let us now denote by $A$ and $B$ the  complex functions
$A = A_{1} + i\,A_{2}$, $B = B_{1} + i\,B_{2} \,,
$
with  $A_{j}$ and $B_{j}$, $j=1,2$,  given by
$$
 A_{1} =  \frac{a^2-b^2}{a^2+b^2}   \,,{\qquad}  A_{2} = \frac{2 a b}{a^2+b^2} \,,    
$$
and 
$$
 B_{1} =   \frac{(a p_b - b p_a)^2}{a^2+b^2} + k_1
 = \frac{J^2}{a^2+b^2} + k_1\,,{\qquad}
 B_{2} =  \frac{J\,(a p_a + b p_b)}{a^2+b^2}\, + k_3 a - k_2 b \,.
$$ 
Then we have the following property : The time-derivative (Poisson bracket with $H_{K2}$) of the function $A$ is proportional to itself and and this property is also true for the function $B$
$$
 \frac{d}{d t}\,A = \{A\,, H_{K2}\} =  2\, i\,\lambda\,A \,,{\qquad}
 \frac{d}{d t}\,B = \{B\,, H_{K2}\} =  2\, i\,\lambda\,B \,, 
$$
where the common factor ${\lambda}$ takes the value 
$$
 {\lambda} = \frac{a p_b - b p_a}{(a^2+b^2)^2} = \frac{J}{(a^2+b^2)^2} \,.
$$
Consequently the Poisson bracket of  the complex function $A B^{*}$ with the  Hamiltonian $H_{K2}$ vanishes 
\begin{eqnarray*}   
  \{A \,B^{*} \,, H_{K2}\}  &=&   \{A  \,,H_{K2}\}\,B^{*}   +   A\,\{B^{*} \,,H_{K2}\}    \cr
    &=&   \bigl( i\, 2\, \lambda\,A \bigr)\,B^{*}   +   A\,\bigl( -\,i\,2\, \lambda\,B^*\bigr)  = 0  \,. 
\end{eqnarray*}

The following proposition summarizes this result. 

\begin{proposition}  \label{proposition1}
Let us consider  the following $(k_1,k_2,k_3)$-dependent Kepler-related Hamiltonian  
$$
  H_{K2}  =   ({\smallonehalf})\,\Bigl( \frac{p_a^2 + p_b^2}{a^2+b^2} \Bigr) +
    \Bigl[\,  \frac{k_1}{a^2+b^2} + \frac{k_2\, a }{a^2+b^2} +  \frac{k_3\, b}{a^2+b^2}  \,\Bigr]
$$
Then, the complex function $J_{34}$ defined as 
$$
  J_{34} = A  B^{*} 
$$
is a (complex) constant of the motion. 
\end{proposition}

The complex function $J_{34}$ determines two real first-integrals 
$$
 J_{34} = J_3 + i\, J_4 \,,{\quad}
 \bigl\{J_3\,, H_{K2}\bigr\} =  0 \,,{\quad}
 \bigl\{J_4\,, H_{K2}\bigr\} =   0 \,, 
$$
whose coordinate expressions are just the two components ($R_x$ and $R_y$) of the generalized  Laplace-Runge--Lenz vector  
\begin{eqnarray*}   
 J_3 &=& {\rm Re}(J_{34})  =  J P_2  +  2 \Bigl[\, k_1 \smallonehalf \,\frac{a^2-b^2}{a^2+b^2} - 
 k_2\frac{a b^2}{a^2+b^2} + k_3 \frac{a^2 b}{a^2+b^2}  \,\Bigr]\,,\cr 
 J_4 &=& {\rm Im}(J_{34})  =  J P_1 -  2 \Bigl[\, k_1  \,\frac{a b}{a^2+b^2} +  \smallonehalf \,k_2\frac{b\, (a^2-b^2)}{a^2+b^2} + \smallonehalf \,k_3 \frac{a \,(b^2-a^2)}{a^2+b^2}  \,\Bigr]\,.
\end{eqnarray*}
As it is well known the existence of this conserved vector is one of the main characteristics of the Kepler problem (the standard Laplace-Runge--Lenz vector correspond to $k_1\ne 0$, $k_2=k_3=0$) and the importance of this fact have led to many authors to the study of Kepler-related systems  admitting generalizations of the Laplace-Runge-Lenz vector \cite{HolasMar90}--\cite{Nikitin14}. 
Now we have arrived to a new property: it can also be obtained as a consequence of this complex formalism.

Summarizing,  
(i) The superintegrability of the Kepler-related Hamiltonian $H_{K2}$  is directly related with the existence of two complex functions, $A$ and $B$,  whose Poisson brackets with the Hamiltonian are proportional, with a common complex factor $2\, i\,\lambda$, to themselves, and (ii),  
The two components of the  $(k_1,k_2,k_3)$-dependent Laplace-Runge-Lenz vector, $J_3$ and $J_4$, appear as the real and imaginary parts of the complex first-integral of motion. 
Remark that $A$ is a complex function of constant modulus one, while the modulus of $B$ is a polynomial of degree four in the momenta given by 
$$
 B\,B^*   =  J_3^2 + J_4^2 = 2 J^2 H_{K2} + 2 J (k_3 p_a - k_2 p_b) + k_1^2 + (k_2 b - k_3 a)^2\,.
$$

Let us now denote by $Y_{34}$ the (complex) Hamiltonian vector field of $J_{34}$
$$
 i(Y_{34})\,\omega_0 = d J_{34}  \,,  
$$
that obviously satisfies $Y_{34}(H_{K2}) = \{H_{K2}\,,J_{34}\} = 0$, 
and by $Y_{A}$ and $Y_{B}$ the Hamiltonian vector fields of $A$ and $B$:
$$
 i(Y_{A})\,\omega_0 = d A  \,,{\quad}  i(Y_{B})\,\omega_0 = d B \,, 
$$
that is 
$$
  Y_{A} = \Bigl(\fracpd{A}{p_a}\Bigr)\fracpd{}{a} + \Bigl(\fracpd{A}{p_b}\Bigr)\fracpd{}{b} - \Bigl(\fracpd{A}{a}\Bigr)\fracpd{}{p_a} - \Bigl(\fracpd{A}{b}\Bigr)\fracpd{}{p_b} 
$$
$$
  Y_{B} = \Bigl(\fracpd{B}{p_a}\Bigr)\fracpd{}{a}+ \Bigl(\fracpd{B}{p_b}\Bigr)\fracpd{}{b}- \Bigl(\fracpd{B}{a}\Bigr)\fracpd{}{p_a}- \Bigl(\fracpd{B}{b}\Bigr)\fracpd{}{p_b} 
$$
Their local coordinate expressions  are, respectively,  given by 
$$
  Y_{A}  = \frac{2}{(a^2+b^2)^2}  \biggl( - 2 a b +i\, (a^2-b^2) \Bigr) 
  \Bigl(b\fracpd{}{p_a} - a \fracpd{}{p_b}\Bigr)
$$
and
$$
  Y_{B} = Y_{Bh}  -  Y_{Bv}
  $$
$$
  Y_{Bh} = \frac{1}{(a^2+b^2)}  \biggl(  2 J \bigl(a\fracpd{}{b} - b \fracpd{}{a} \bigr) + i\,
  \Bigl(  J  \bigl(a\fracpd{}{a} + b \fracpd{}{b} \bigr) + (a p_a + b p_b) \bigl(a\fracpd{}{b} - b \fracpd{}{a} \bigr)  \biggl)
$$
$$
  Y_{Bv} = \frac{1}{(a^2+b^2)^2} W \Bigl(b\fracpd{}{p_a} - a \fracpd{}{p_b}\Bigr) + i\,\Bigl(k_3\fracpd{}{p_b} - k_2 \fracpd{}{p_a}\Bigr)
  $$
  where $W$ denotes the following complex function 
  $$
  W = 2 \Bigl((a^2-b^2) p_a p_b - a b (p_a^2-p_b^2) \Bigr)+ i\,\Bigl( (a^2-b^2)(p_a^2-p_b^2)+ 4 a b p_a p_b\Bigr)
  = i\,(a + i b)^2(p_a-ip_b)^2
  $$ 
Then, the vector field $Y_{34}$  appears as a linear combination of $Y_{A}$ and $Y_B^*$; more specifically we have 
$$
 Y_{34} = B^*\,Y_{A} + A\,Y_{B}^* = Y + Y' \,,{\quad} 
 Y = B^*\,Y_A   \,,{\quad}   Y' = A\,Y_B^* \,. 
$$
The vector field $Y_{34}$ is certainly a symmetry of the Hamiltonian system $(T^*Q,\omega_0,H_{K2})$, 
but the two vector fields, $Y$ and $Y'$, are neither symmetries of the symplectic form $\omega_0$ 
(that is,  $\CaL_{Y}\,\om_0\ne 0$ and $\CaL_{Y'}\om_0\ne 0$) nor symmetries of the Hamiltonian (that is, $\CaL_{Y}H_{K2}\ne 0$ and $\CaL_{Y'}H_{K2}\ne 0$). 
Moreover, remark that they are not symmetries of the dynamics, because 
$$
 [Y,\Ga_{K2}]  \ne 0 \,,{\quad} [Y',\Ga_{K2}]  \ne 0  \,,{\quad} i(\Ga_{K2})\,\omega_0 = d H_{K2}\,. 
$$
Then it can be proved (by direct computation)
that the Lie bracket of  the dynamical vector field $\Gamma_{K2}$ with $Y$ is given by 
$$
 [\Ga_{K2}, Y]  =  i\,J_{34} X_{\lambda},
$$
where $X_{\lambda}$ is the Hamiltonian vector field of the function $\lambda$.  
The vector field  $X_{\lambda}$ on the right hand side represents an obstruction for $Y$ to be a dynamical symmetry. Only when  ${\lambda}$ be a numerical constant  the vector field  $Y$ (and also $Y'$) is a dynamical symmetry of $\Gamma_K$. 

In the following   $\Omega$ will denote   the complex 2-form defined as 
$$
 \Omega = dA\,\wedge\,dB^*  \,. 
$$
The two complex 2-forms $\om_Y$ and $\om_Y'$  obtained by Lie derivative of $\om_0$, i.e. 
$$
 \CaL_{Y}\,\om_0  = \om_Y \,,{\quad} \CaL_{Y'}\,\om_0  = \om_Y'  \,, 
$$ 
are such 
$$
 \CaL_{Y}\,\om_0  = i_Y (d\om_0) + d(i_Y\om_0) = d(i_Y\om_0) = d\bigl(B^*\,dA\bigr)  
 =-\,\Omega 
$$
$$
 \CaL_{Y'}\,\om_0  = i_{Y'} (d\om_0) + d(i_{Y'}\om_0) = d(i_{Y'}\om_0) = d\bigl(A\,dB^*\bigr)  = \Omega 
$$

Using  the preceding results we can prove: 
\begin{proposition}  \label{proposition2}
The Hamiltonian vector field $\Ga_{K2}$ of the $(k_1,k_2,k_3)$-dependent Kepler-related  problem $H_{K2}$ is a quasi-Hamiltonian system with respect to the complex 2-form $\Omega$.  
\end{proposition}
{\sl Proof}.-   The  contraction of the vector field $ \Ga_{K2}$ with the complex 2-form $\Omega$ gives:
 $$
 i(\Ga_{K2})\,\Omega =  \Ga_{K2} (A)\,dB^* - \Ga_{K2}(B^*)\,dA \,, 
$$
and  recalling that 
$$
  \Ga_{K2} (A) = \{A\,, H_{K2}\} = i\,\lambda\,A    \,,{\quad}
  \Ga_{K2} (B^*) = \{B^*\,,H_{K2}\} = -\, i\,\lambda\,B^* \,, 
$$
we arrive to 
$$
  i(\Ga_{K2})\,\Omega = ( i\,\lambda\,A) \,dB^* + (i\,\lambda\,B^*)\,dA
  = i\,\lambda\,d(A B^*)  \,. 
$$
\hfill{$\square$}

The complex 2-form $\Omega$ can be written as
$$
 \Omega = \Omega_1 + i\,\Omega_2
 $$
 where the two real 2-forms, $\Omega_1 = {\rm Re}(\Omega)$ and $\Omega_2={\rm Im}(\Omega)$, 
 take the form                
\begin{eqnarray*}   
  \Omega_1 &=& d A_{1} \,\wedge\, dB_{1} + d A_{2} \,\wedge\, dB_{2}  \cr
  &=& \al_{12}\,  da \,\wedge\, db  + \al_{13}\, da \,\wedge\,dp_a  
  + \al_{14}\, da\,\wedge\,dp_b  + \al_{23}\,  db \,\wedge\, dp_a  
  + \al_{24}\, db \,\wedge\,dp_b  \cr
  \Omega_2 &=& -\,d A_{1} \,\wedge\, dB_{2} + d A_{2} \,\wedge\, dB_{1}   \cr
 &=& \be_{12}\,  da \,\wedge\, db  + \be_{13}\, da \,\wedge\,dp_a  
  + \be_{14}\, da\,\wedge\,dp_b   + \be_{23}\,  db \,\wedge\, dp_a  
  + \be_{24}\, db \,\wedge\,dp_b  
\end{eqnarray*}   
with $\alpha_{ij}$ and $\beta_{ij}$ being given by 
$$
  \al_{12} = \frac{2}{(a^2+b^2)^2}\Bigl(  (a^2-b^2)(b k_2 - a k_3)\Bigr)    \,,{\quad} 
  \al_{13} = \frac{2b}{(a^2+b^2)^2}\Bigl(  2 a b p_a - (a^2+b^2) p_b\Bigr)   \,, 
$$
$$
  \al_{14} = \frac{2b}{(a^2+b^2)^2}\Bigl(  2 a b p_b - (a^2+b^2) p_a\Bigr)    \,,{\quad} 
  \al_{23} = \frac{2a}{(a^2+b^2)^2}\Bigl(  2 a b p_a - (a^2+b^2) p_b\Bigr)   \,,
$$
$$
  \al_{24} = \frac{2a}{(a^2+b^2)^2}\Bigl(  2 a b p_b - (a^2+b^2) p_a\Bigr)  \,,{\quad}
  \al_{34} =   0    \,. 
$$
and
$$
  \beta_{12} = \frac{4 a b (a k_3 - b k_2)}{(a^2+b^2)^2}  \,,{\quad}
  \beta_{13} =   \frac{4 b^3 p_a}{(a^2+b^2)^2}    \,,{\quad}
  \beta_{14} =   -\, \frac{4 b^2 b p_b}{(a^2+b^2)^2}    \,. 
$$    
$$
  \beta_{23} =  -\, \frac{4 b b^2 p_a}{(a^2+b^2)^2}   \,,{\quad}
  \beta_{24} =   \frac{4 a^3 p_b}{(a^2+b^2)^2}    \,,{\quad}
  \beta_{34} =   0    \,. 
$$    
Then we have  
$$
  i(\Ga_{K2})\,\Omega_1 = -\,\lambda\,dJ_4 \,,{\quad} 
  i(\Ga_{K2})\,\Omega_2 =  \lambda\,dJ_3  \,, 
$$
what means that $\Ga_{K2}$ is also  quasi-bi-Hamiltonian with respect to the two real 2-forms $(\om_0,\Omega_1)$ or $(\om_0,\Omega_2)$. 

Therefore, the two complex functions, $A$ and $B$,  that  determine the existence of superintegrability (existence of additional constants of motion) are also directly related with the existence of  quasi-bi-Hamiltonian structures [first complex $(\om_0,\Omega)$ and then real $(\om_0,\Omega_1,\Omega_2)$].     

Remark that the complex 2-form $\Omega$ is well defined  but it is not symplectic. 
In fact, from the above expressions in coordinates we have 
$\Omega_1\,\wedge\, \Omega_1 = 0$,
$\Omega_2\,\wedge\, \Omega_2 = 0$, and 
$\Omega_1\,\wedge\, \Omega_2 = 0$,
and therefore we obtain 
$$
  \Omega\,\wedge\, \Omega  =  \bigl(\Omega_1\,\wedge\, \Omega_1 - \Omega_2\,\wedge\, \Omega_2\bigr) + 2\, i \,\Omega_1\,\wedge\, \Omega_2 = 0   \,. 
$$

The distribution defined by the kernel of  $\Omega_1$, that is two-dimensional,  is given by 
$$
 \Ker\,\Omega_1 = \{\,f_1\,X_{11}+ f_2\,X_{12} \ |\ f_1,f_2 : \mathbb{R}^2\times\mathbb{R}^2\to\mathbb{C} \,\},
$$
where the vector fields $X_{11}$ and $X_{12}$ are 
$$
  X_{11}= \Bigl(  2 a b p_b - (a^2+b^2) p_a\Bigr) \fracpd{}{p_a} 
  + \Bigl(  2 a b p_b - (a^2+b^2) p_a\Bigr)  \fracpd{}{p_b} \,, 
$$
$$
X_{12}= a \fracpd{}{p_a} + b \fracpd{}{b} - \frac{(a^2-b^2)(a k_3 - b k_2)}{(a^2+b^2) p_a - 2 a b p_a} \fracpd{}{p_b} \,. 
$$
In a similar way the kernel of $\Omega_2$ is given by 
$$
 \Ker\,\Omega_2 = \{\,f_1\,X_{21}+ f_2\,X_{22} \ |\ f_1,f_2 : \mathbb{R}^2\times\mathbb{R}^2\to\mathbb{C} \,\},
$$
where the vector fields $X_{21}$ and $X_{22}$ are 
$$
  X_{21}= a^2 p_b \fracpd{}{p_a}   + b^2 p_a \fracpd{}{p_b}  
\,,{\qquad} 
X_{22}= a \fracpd{}{p_a} + b \fracpd{}{b} - \frac{b(a k_3 - b k_2)}{a p_b} \fracpd{}{p_b} \,. 
$$
We have
$$
 [\Ker\,\Omega_1\,,\Ker\,\Omega_1] \subset  \Ker\,\Omega_1\,,{\quad}
[\Ker\,\Omega_2\,,\Ker\,\Omega_2] \subset  \Ker\,\Omega_2\,.
$$

If  $Y_3$ and $Y_4$ are the Hamiltonian vector fields 
(with respect to the canonical symplectic form $\om_0$)  
of the first integrals $J_3$ and $J_4$, then the  dynamical vector field $\Ga_{K2}$ is orthogonal to $Y_4$ 
with respect to the structure $\Omega_1$ and it is also orthogonal to $Y_3$ with respect to the structure $\Omega_2$, 
that is, 
$$
 i(\Ga_{K2})\,i(Y_4)\,\Omega_1 = 0 \,, {\qquad}  i(\Ga_{K2})\,i(Y_3)\,\Omega_2 = 0 \,. 
$$

The bi-Hamiltonian structure $(\om_0,\Omega)$ determines a complex recursion operator $
R$ defined as 
$$
  \Omega(X,Y) = \om_0(RX,Y) \,,{\quad} \forall X,Y \in \mathfrak{X}(T^*Q) \,. 
$$
But as $\Omega$ and $R$ are complex, we can introduce two  real recursion operator 
$R_{1}$ and $R_{2}$ defined as 
$$
  \Omega_{1}(X,Y) = \om_0(R_{1}X,Y)  \,,{\qquad}   \Omega_{2}(X,Y) = \om_0(R_{2}X,Y)  \,. 
$$
We recall that $\wh{\om_0}$ is the map $\wh{\om_0}:\mathfrak{X}(T^*Q)\to\wedge^1(T^*Q)$ 
given by contraction, that is $\wh{\om_0}(X) = i(X)\om_0$, and then the  nondegenerate character 
of $\om_0$ means that the map  $\wh{\om_0}$ is a bijection. 
Using this notation we can write the two operators $R_{1}$ and $R_{2}$ as follows    
$$
  R_{1} = \wh{\om_0}^{-1} \circ \wh{\Omega_{1}}   \,,{\qquad}
  R_{2} = \wh{\om_0}^{-1} \circ \wh{\Omega_{2}}  \,.
$$  
Then we have the following properties 
\begin{itemize}
\item  [(i)] The coordinates expressions of $R_{1}$  and $R_{2}$  are
\begin{eqnarray*}   
 R_{1} &=& \Bigl[\al_{13} \fracpd{}{a} + \al_{14} \fracpd{}{b} - \al_{12} \fracpd{}{p_b}\Bigr] \otimes da
 + \Bigl[\al_{23} \fracpd{}{a} + \al_{24}\fracpd{}{b}+ \al_{12}\fracpd{}{p_a}\Bigr] \otimes db
 \cr&+& \Bigl[\al_{13} \fracpd{}{p_a} + \al_{23}\fracpd{}{p_b}\Bigr] \otimes dp_a
  + \Bigl[\al_{14} \fracpd{}{p_a} + \al_{24}\fracpd{}{p_b}\Bigr] \otimes dp_b
\end{eqnarray*}   
and 
\begin{eqnarray*}   
 R_{2} &=& \Bigl[\be_{13} \fracpd{}{a} + \be_{14} \fracpd{}{b} - \be_{12} \fracpd{}{p_b}\Bigr] \otimes da
 + \Bigl[\be_{23} \fracpd{}{a} + \be_{24}\fracpd{}{b}+ \be_{12}\fracpd{}{p_a}\Bigr] \otimes db
 \cr 
 &+& \Bigl[\be_{13} \fracpd{}{p_a} + \be_{23}\fracpd{}{p_b}\Bigr] \otimes dp_a
  + \Bigl[\be_{14} \fracpd{}{p_a} + \be_{24}\fracpd{}{p_b}\Bigr] \otimes dp_b
\end{eqnarray*}   
\item [(ii)] $R_{1}$  and $R_{2}$ have two different eigenvalues doubly degenerate and one of them is null (that is, $\lambda_1=\lambda_2=0$, $\lambda_3=\lambda_4\ne 0$). Therefore we have 
$$
  \det [R_{1}] =\det [R_{2}] = 0\,, 
$$
what is a consequence of the singular character of $\Omega_1$ and $\Omega_2$. 
 \end{itemize}

\section{Hamiltonian $H_{K2}$.  New complex functions and new quasi-bi-Hamiltonian structures} 

The expressions of the two complex functions $A$ and $B$ (studied in the previous sections (\ref{Sec2})) have a rather different form (lack of symmetry between these functions). Now, in this new Section we present a new approach that makes use of two new complex functions (to be denoted by $M_a$ and $M_b$) that are quite similar one to the other 
(it is a more symmetric approach that generalizes results obtained in \cite{CarRan16Sigma}).

Let us now consider a second set of  complex functions functions 
$M_a = M_{a1} + i\,M_{a2}$, $ M_b = M_{b1} + i\,M_{b2}$, 
with   $M_{a j}$ and $M_{b j}$,  $j=1,2$,  defined by: 
$$
 M_{a1} = \frac{1}{\sqrt{a^2+b^2}} \Bigl( {J p_a} -  {(k_2 b - k_3 a)a}  \Bigr) \,,{\qquad}
 M_{a2} = \frac{1}{\sqrt{a^2+b^2}} \Bigl( {-J p_b } -2 k_1 a + (k_2 b - k_3 a)b  \Bigr)  \,,
$$
 and 
$$
 M_{b1} = \frac{1}{\sqrt{a^2+b^2}} \Bigl( {J p_b} -  {(k_2 b - k_3 a)b}  \Bigr) \,,{\qquad}
 M_{b2} = \frac{1}{\sqrt{a^2+b^2}} \Bigl( {J p_a } -2 k_1 b - (k_2 b - k_3 a) a \Bigr)  \,. 
$$
Then we have the following property $$
   \{M_a\,,\,H_{K2}\} = i\,\lambda\,M_{a}   \,,{\quad} 
   \{M_b\,,\,H_{K2}\} = i\,\lambda\,M_{b}  \,. 
$$

\begin{proposition}  \label{proposition3}
The complex function $K_{34}$ defined as 
$$
  K_{34} = M_a \,M_b^{*} 
$$
is a (complex) constant of the motion for the dynamics of the  $(k_1,k_2,k_3)$-dependent Kepler-related  system  described by the Hamiltonian $H_{K2}$. 
\end{proposition}
The proof  is quite similar to the proof of the previous Proposition  \ref{proposition1}.

Note that the modulus of the complex functions $M_a$ and  $M_b$, that are constants of motion,  are given by 
\begin{eqnarray*}   
 M_a\,M_a^* &=& 2\,\bigl( J^2 H_{K2} + k_1 R_x + J (k_3 p_a - k_2 p_b) ) + (k_2 b  - k_3 a )+ 2 k_1^2  \,,  \cr
 M_b\,M_b^* &=& 2\,\bigl( J^2 H_{K2} - k_1 R_x  + J (k_3 p_a - k_2 p_b) ) + (k_2 b  - k_3 a )+ 2 k_1^2  \,. 
 \end{eqnarray*}   
The complex function $K_{34}$ determines two real functions that are first integrals for the $H_{K2}$:
$$
 K_{34} = K_3 + i\, K_4 \,,{\quad}
 \bigl\{K_3\,, H_{K2}\bigr\} =  0 \,,{\quad}
 \bigl\{K_4\,, H_{K2}\bigr\} =  0 \,,
$$
with $K_3$ and $K_4$ given by 
\begin{eqnarray*}   
K_3 &=& {\rm Re}(K_{34})   = M_{a1}M_{b1} + M_{a2}M_{b2} 
  =   J P_1 - \Bigl( \frac{2 k_1a b}{a^2+b^2} +   \frac{(k_2 b-k_3 a) (a^2 - b^2)}{(a^2 + b^2)} \Bigr)   \,,  \cr 
K_4 &=&  {\rm Im}(K_{34}) =  M_{a2}M_{b1} - M_{a1}M_{b2} 
 = 2 J^2 H_{K2} + 2 J (k_3 p_a - k_2 p_b)  + (k_2 b - k_3 a)^2   \,.
\end{eqnarray*}   
The function $K_3$ is the component $R_y$ of the generalized Laplace-Runge-Lenz constant,   
$K_4$  is a  fourth order  in the momenta polynomial and $ M_a\,M_a^*  -  M_b\,M_b^* $ is just the other component $R_x$ of the above mentioned vector
$$
 M_a\,M_a^*  -  M_b\,M_b^*  = 4 k_1 \Bigl[ J P_2  +  2 \Bigl(\, k_1 \smallonehalf \,\frac{a^2-b^2}{a^2+b^2} - 
 k_2\frac{a b^2}{a^2+b^2} + k_3 \frac{a^2 b}{a^2+b^2}  \Bigr)\,\Bigr]  \,.
$$

Let us now denote by $Z_{34}$ the Hamiltonian vector field of the function $K_{34}$, i.e. 
$ i(Z_{34})\,\omega_0 = d Z_{34} $, such that $Z_{34}(H_{K2}) = 0$,   
and by $Z_{a}$ and $Z_{b}$ the Hamiltonian vector fields of the complex functions $M_a$ and $M_b$, that is,  
$$
 i(Z_{a})\,\omega_0 = d M_{a}  \,,{\quad}   i(Z_{b})\,\omega_0 = d M_{b} \,. 
$$
Their  coordinate expressions  are given by  
$$
  Z_a = \Bigl(\fracpd{M_a}{p_a}\Bigr)\fracpd{}{a} + \Bigl(\fracpd{M_a}{p_b}\Bigr)\fracpd{}{b} - \Bigl(\fracpd{M_a}{a}\Bigr)\fracpd{}{p_a} - \Bigl(\fracpd{M_a}{b}\Bigr)\fracpd{}{p_b} 
  = Z_{a0} + k_1 Z_{a1} + k_2 Z_{a2} + k_3 Z_{a3}
$$
with $Z_{a0}$ and $Z_{ai}$, $i=1,2,3$,  given by 
$$
  Z_{a0} = \frac{1}{\sqrt{a^2+b^2}}  \biggl(  
  \bigl(\bigl(a p_b - 2 b p_a + i\,b p_b\bigr)  \fracpd{}{a}   + \bigl( a p_a + i\,(b p_a -2 a p_b  )\bigr)  \fracpd{}{b}   -   \frac{(a p_a + b p_b)}{a^2+b^2}   \Bigl( (p_a-i\,p_b)\,  (b \fracpd{}{p_a}  - a \fracpd{}{p_b} )\Bigr) \biggr),
$$
\begin{eqnarray*}   
 Z_{a1} &=&    \frac{2 i\,b}{(a^2+b^2)^{3/2}}  \bigl(b \fracpd{}{p_a}  - a \fracpd{}{p_b} \bigr)\,,  \cr 
 Z_{a2} &=&     \frac{1}{(a^2+b^2)^{3/2}} \Bigl[ \bigl(b^3 \fracpd{}{p_a}  + a^3 \fracpd{}{p_b} \bigr) 
  + i\,b\Bigl( a b  \fracpd{}{p_a} - (2 a^2 + b^2)\ \fracpd{}{p_b} \bigr) \Bigr] \,,  \cr 
 Z_{a3} &=&      \frac{1}{(a^2+b^2)^{3/2}} \Bigl[  a\Bigl(- (a^2 + 2 b^2) \fracpd{}{p_a} + a b  \ \fracpd{}{p_b}\Bigr)
  + i\, \bigl(b^3 \fracpd{}{p_a}  + a^3 \fracpd{}{p_b} \bigr)  \bigr) \Bigr] \,,  
\end{eqnarray*}   

and 
$$
  Z_b = \Bigl(\fracpd{M_b}{p_a}\Bigr)\fracpd{}{a}+ \Bigl(\fracpd{M_b}{p_b}\Bigr)\fracpd{}{b}- \Bigl(\fracpd{M_b}{a}\Bigr)\fracpd{}{p_a}- \Bigl(\fracpd{M_b}{b}\Bigr)\fracpd{}{p_b} 
   =  Z_{b0} + k_1 Z_{b1} + k_2 Z_{b2} + k_3 Z_{b3}
$$
with $Z_{b0}$ and $Z_{bi}$, $i=1,2,3$,  given by 
$$
  Z_{b0} = \frac{1}{\sqrt{a^2+b^2}}  \biggl(   \bigl(- b p_b + i\,(a p_b - 2 b pa)  \bigr) \fracpd{}{a} 
    + \bigl( 2 a p_b - b pa + i\,a p_a  )\bigr)  \fracpd{}{b}   -   \frac{(a p_a + b p_b)}{a^2+b^2}   \Bigl( (p_b+ i\,p_a)\,  (b \fracpd{}{p_a}  - a \fracpd{}{p_b} )\Bigr) \biggr),
$$
\begin{eqnarray*}   
 Z_{b1} &=&    -\frac{2 i\,a}{(a^2+b^2)^{3/2}}  \bigl(b \fracpd{}{p_a}  - a \fracpd{}{p_b} \bigr)\,,  \cr 
 Z_{b2} &=&     \frac{1}{(a^2+b^2)^{3/2}} \Bigl[ b\Bigl(- a b  \fracpd{}{p_a} + (2 a^2 + b^2)\ \fracpd{}{p_b} \bigr)  +
  i\,\bigl(b^3 \fracpd{}{p_a}  + a^3 \fracpd{}{p_b} \bigr) 
   \Bigr] \,,  \cr 
 Z_{b3} &=&    -  \frac{1}{(a^2+b^2)^{3/2}} \Bigl[ \bigl(b^3 \fracpd{}{p_a}  + a^3 \fracpd{}{p_b} \bigr) + i\, a\Bigl(
    (a^2 + 2 b^2) \fracpd{}{p_a} - a b   \fracpd{}{p_b}
     \bigr) \Bigr]    \,. 
\end{eqnarray*}   

Now recalling that 
$$
 dZ_{34} = d \,\bigl(M_a M_b^*\bigr) = M_b^*\,d \bigl(M_a\bigr)  + M_a\, d \bigl(M_b^*\bigr),
$$
we obtain 
$$
 Z_{34} = M_b^*\,Z_a + M_a\,Z_b^* = Z + Z'\,,{\quad} \textrm{where}\quad 
 Z = M_b^*\,Z_a   \,,{\quad}   Z' = M_a\,Z_b^* \,. 
$$

In the following we will denote by $\Omega_{M}$ the complex 2-form defined as $\Omega_{M} = dM_a\,\wedge\,dM_b^*$. 
Then the two 2-forms $\om_Z$ and $\om_Z'$  obtained by Lie derivation of $\om_0$ with respect to $Z$ and $Z'$ are given by 
$$
 \CaL_{Z}\,\om_0  = \om_Z  = -\,\Omega_{M}   \,,{\qquad} 
 \CaL_{Z'}\,\om_0 = \om_Z' = \Omega_{M}.
$$ 

\begin{proposition}  \label{proposition4}
The Hamiltonian vector field $\Ga_{K2}$ of the $(k_1,k_2,k_3)$-dependent Kepler-related  problem $H_{K2}$ is a quasi-Hamiltonian system with respect to the complex 2-form $\Omega_{M}$.  
\end{proposition} 
{{\sl Proof}.-} This can be proved by a direct computation:
  $$\begin{array}{rcl}
 i(\Ga_{K2})\,\Omega_{M} &=&    \Ga_{K2} (M_a)\,dM_b^* - \Ga_{K2}(M_b^*)\,dM_a  \\
   &=& ( i\,\lambda\,M_{a}) \,dM_b^* + (i\,\lambda\,M_b^*)\,dM_a
  = i\,\lambda\,d(M_a M_b^*).   
\end{array}   
$$
 \hfill${\square}$
 
 The complex 2-form $ \Omega_M$ can be decomposed as
$$
 \Omega_M= \Omega_{M1} + i\,\Omega_{M2},
 $$
 where the two real 2-forms,  $\Omega_{M1} = {\rm Re}(\Omega_M)$ and 
 $\Omega_{M2}={\rm Im}(\Omega'_M$,  take the form 
$$
  \Omega_{M1} = d M_{a1} \,\wedge\, dM_{b1} + d M_{a2} \,\wedge\, dM_{b2} \,,{\qquad} 
  \Omega_{M2} = -\,d M_{a1} \,\wedge\, dM_{b2} + d M_{a2} \,\wedge\, dM_{b1} \,,
$$
and then considering the real and imaginary parts we obtain: 
$$
  i(\Ga_{K2})\,\Omega_{M1} = -\,\lambda\,dK_4 \,,{\quad} 
  i(\Ga_{K2})\,\Omega_{M2} =  \lambda\,dK_3  \,, 
$$
what means that $\Ga_{K2}$ is also quasi-bi-Hamiltonian with respect to the two real 2-forms $(\om_0, \Omega_{M1})$ and $(\om_0, \Omega_{M2})$.

The coordinate expressions of   $\Omega_{M1}$ and $  \Omega_{M2}$ are  
$$
  \Omega_{M1} = \ \frac{2 }{(a^2+b^2)^2}   \Bigl( \al_{12} da\,\wedge\,db   + \al_{13} da\,\wedge\,dp_a + \al_{14} da\,\wedge\,dp_b + \al_{23} db\,\wedge\,dp_a + \al_{24} db\,\wedge\,dp_b + \al_{34} dp_a\,\wedge\,dp_b ) \Bigr) ,
$$
$$
  \Omega_{M2} = \ \frac{2 k_1}{(a^2+b^2)^2} 
    \Bigl(\,  \be_{12} da\,\wedge\,db + \be_{13} da\,\wedge\,dp_a + \be_{14} da\,\wedge\,dp_b + \be_{23} db\,\wedge\,dp_a + \be_{24} db\,\wedge\,dp_b   \Bigr)  ,
$$
with $\alpha_{ij}  = \alpha_{ijk_1} + \alpha_{ijk} $ and $\beta_{ij}$  given by 
$$ \begin{array}{rcl rcl}
 \al_{12k_1} &=& 0  \,,  &\quad
 \al_{13k_1} &=& -\,J\bigl( (a p_a + b p_b) p_b + 2 k_1 b \bigr) b    \,,      \cr
 \al_{14k_1} &=& J\bigl(  (a p_a + b  p_b) p_a  + 2 k_1 a  \bigr) b  \,,    &\quad
 \al_{23k_1} &=& J\bigl(  (a p_a  + b p_b) p_b  + 2 k_1 b  \bigr) a \,,   \cr
\al_{24k_1} &=& -\,J\bigl(  (a p_a + b p_b) p_a + 2 k_1 a \bigr) a \,,  &\quad
\al_{34k_1} &=& 2 J^2 (a^2+b^2) \,, 
 \end{array}  
 $$
\begin{eqnarray*}   
 \al_{12k} &=&    (k_2  b-  k_3 a) \bigl( ( k_2 b  -  k_3 a)  (a^2 + b^2) +  J (a p_a + b p_b) \bigr) \cr
 \al_{13k} &=&   k_2  b^2 (2 a b p_a - a^2 p_b + b^2 p_b) + k_3  b  (2 b^3 p_a - a^3 p_b - 3 a b^2 p_b) \cr
  \al_{14k} &=&  -\, k_2 b^2  (a^2 p_a - b^2 p_a + 2 a b p_b) +  k_3  a  (-a^2 b p_a - 3 b^3 p_a + 2 a^3 p_b + 4 a b^2 p_b)  \cr
   \al_{23k} &=& k_2  b  (-4 a^2 b p_a - 2 b^3 p_a + 3 a^3 p_b + a b^2 p_b)   -\, k_3 a^2  (-2 a b p_a + a^2 p_b - b^2 p_b)   \cr
   \al_{24k} &=&   -\, k_2 a  (-3 a^2 b p_a - b^3 p_a + 2 a^3 p_b) -  k_3 a^2  (a^2 p_a - b^2 p_a + 2 a b p_b)  \cr
    \al_{34k} &=& 0 \,,
\end{eqnarray*}   
 and   
 $$ \begin{array}{rcl rcl}
 \be_{12} &=&  -\, (a^2-b^2)(k_2 b - k_3 a)      \,, &\quad
 \be_{13} &=& \bigl( (a^2 p_b + b^2) p_b - 2 a b p_a   \bigr) b  \,,  \cr
 \be_{14} &=& \bigl( (a^2 p_b + b^2) p_a - 2 a b p_b   \bigr) b    \,,  &\quad
 \be_{23} &=&  -\, \bigl( (a^2 p_b + b^2) p_b - 2 a b p_a   \bigr) a   \,, \cr
\be_{24} &=&  -\, \bigl( (a^2 p_b + b^2) p_a - 2 a b p_b   \bigr) a  \,, &\quad
\be_{34} &=& 0  \,. 
 \end{array}  
 $$

We close this section with the following  properties:
\begin{itemize}
\item[(i)]  The two real 2-forms are not symplectic. In fact we have verified that 
$\Omega_{M1}\,\wedge\,\Omega_{M1} = 0$,
$\Omega_{M2}\,\wedge\,\Omega_{M2} = 0$, and also 
$\Omega_{M1}\,\wedge\,\Omega_{M2} = 0$. 
 
 \item[(ii)]  These two 2-forms, $\Omega_{M1}$ and $\Omega_{M2}$, determine two recursion operators 
($(1,1)$ tensor fields) $R'_{1}$ and $R'_{2}$ defined as 
$$
  \Omega_{M1}(X,Y) = \om_0(R'_{1}X,Y)  \,,{\qquad}   \Omega_{M2}(X,Y) = \om_0(R'_{2}X,Y)  \,, 
$$
or in an equivalent way  
$$
  R'_{1} = \wh{\om_0}^{-1} \circ \wh{\Omega_{M1}}   \,,{\qquad}
  R'_{2} = \wh{\om_0}^{-1} \circ \wh{\Omega_{M2}}  \,. 
$$  
As in the section (\ref{Sec2}), a consequence of the singular character of  $\Omega_{M1}$ and $\Omega_{M2}$ is that 
$$
  \det [R'_{1}] =\det [R'_{2}] = 0\,.   
$$   

\item[(iii)]
If we denote by $Z_3$ and $Z_4$ the Hamiltonian vector fields (with respect to the canonical 
symplectic form $\om_0$)  of the integrals $K_3$ and $K_4$, then the  dynamical vector field 
$\Ga_{K2}$ is orthogonal to $Z_4$ with respect to the structure $\Omega_{M1}$ and it is also orthogonal 
to $Z_3$ with respect to the structure $\Omega_{M2}$, that is, 
$$
 i(\Ga_{K2})\,i(Z_4)\,\Omega_{M1} = 0 \,, {\qquad}  i(\Ga_{K2})\,i(Z_3)\,\Omega_{M2} = 0 \,. 
$$
\end{itemize}

\section{Final comments } 

We have proved that certain geometric properties (previously studied in \cite{CarRan16Sigma})  characterizing the super\-integrability of the standard Kepler  problem ($k_1\ne 0$, $k_2=k_3=0$) can be generalized (introducing the appropriate changes) to the more general  $(k_1,k_2,k_3)$-dependent Kepler-related problem $H_{K2}$.  In fact,  we have proved that the super\-integrability  of this more general Hamiltonian is related with the Poisson bracket properties of certain complex functions   (we have presented two different approaches)  and also that these functions are directly related with the existence of  quasi-bi-Hamiltonian structures. 

We close pointing out some open questions.  
The complex functions method presented in this paper (as well in some other previous papers mentioned in the Introduction) is restricted to the two dimensional case;  it is convenient to study the generalization to the three-dimensional case (the multiple separability of three-dimensional systems was first studied in \cite{Ev90Pra})  and  also to constant curvature spaces (the superintegrability of some particular systems was studied in \cite{Ran14JPa,Ran15PLa} making use of curvature-dependent polar coordinates); the generalization of the system studied in this paper must be done making use of curvature-dependent parabolic coordinates. 

Finally, the complex functions ($A,B$) or ($M_a,M_b$)  are important for two reasons since they determine the integrals of motion ($A  B^{*}$ or $M_a M_b^{*}$)  and also the geometric structures; probably there are some additional properties hidden behind these functions deserving be studied making use of tools of complex differential geometry.

\section*{Acknowledgments}

This work has been  supported by the research projects PGC2018-098265-B-C31 (MINECO, Madrid)  and DGA-E24/1 (DGA, Zaragoza).

{\small

   }

\end{document}